\documentclass
[aps,prb,endfloat,twocolumn,showpacs,preprintnumbers,amsmath,amssymb]{revtex4}
\usepackage{natbib}
\usepackage[dvips]{graphicx}
\usepackage{wrapfig}
\usepackage{dcolumn}
\usepackage{bm}
\usepackage{times}
\usepackage{mathptmx}
\usepackage{textcomp}
\usepackage[usenames]{color}
\usepackage{ulem}
\def\geqap{\,\raise 2pt \hbox{$>\kern-11pt \lower 5pt \hbox{$\sim$}$}\,}
\def\leqap{\,\raise 2pt \hbox{$<\kern-10pt \lower 5pt \hbox{$\sim$}$}\,}
\begin{document}
%
\title{Numerical study of $t_{2g}$ orbital system with ferromagnetic polarization}

\author{Takayoshi Tanaka$^{\ast}$ and Sumio Ishihara} 
\affiliation{Department of Physics, Tohoku University, Sendai 980-8578, Japan}
\date{\today}
\begin{abstract}

Finite temperature orbital state in a ferromagnetic Mott insulator with triply-degenerate $t_{2g}$ orbital 
is investigated numerically.  We employ the quantum Monte Carlo simulation with the loop algorithm. Indications for conventional staggered-type orbital order are not remarkable down to the lowest temperature to which the present simulation can get access. Physical parameters monitoring the off-diagonal orbital order, 
which is characterized by a linear combination of the $(d_{yz}, d_{zx}, d_{xy})$ orbital-wave functions with equal weights, 
are not conspicuous. It is found that a orbital gap-like behavior appears in the uniform orbital susceptibility. This is supported by a threshold behavior in the staggered correlation function in a calculation with the additional Ising-type interaction. Some rigorous remarks for the long-range orbital order are also presented. 

\end{abstract}
%

\maketitle

Orbital degree of freedom and its strong coupling with spin, charge and lattice 
bring about a variety of fascinating physical phenomena in 
transition-metal oxides.~\cite{maekawa04,tokura00,imada98} 
Colossal magnetoresistance effect observed in vicinity of charge/orbital ordered state in perovskite manganites is a typical example. 
Classical aspects of the orbital, e.g. long range orbital orders accompanied with the Jahn-Teller distortion and their roles on the magnetic exchange interaction, have been fully examined and been almost settled nowadays. 
On the other hand, quantum aspects of the orbital are still unrevealed and are expected to open a window for a new research field in correlated electron system. 

Perovskite titanate $R$TiO$_3$ ($R$: rare-earth metal ion) is one of the candidates as a material where quantum orbital physics is realized. 
A formal valence of Ti ion is 3+, and one $3d$ electron occupies one of the triply degenerate $t_{2g}$ orbitals. 
In this paper, we denote, for simplicity, the three $t_{2g}$ orbitals, ($d_{yz}$, $d_{zx}$, $d_{xy}$) by ($d_\alpha$, $d_{\beta}$, $d_{\gamma}$), respectively. 
Because of a weak Jahn-Teller coupling and the novel symmetry of the $t_{2g}$ orbital-wave functions, 
quantum aspects of the orbital are expected to become evident.~\cite{keimer00, khaliullin00,harris03}

In particular, a ferromagnetic Mott insulator with the $t_{2g}$ orbital degeneracy attracts attention as a simple orbital system where spin and charge degrees of freedom are quenched.~\cite{ishihara02,mochizuki03,ko} 
One representative material is YTiO$_3$, where the 
isotropic spin-wave dispersion relation is observed by the inelastic neutron scattering.~\cite{ulrich02} 
This result seems to contradict a long-range orbital order confirmed by several experiments.~\cite{nakao02,kubota04,akimitsu01,itoh99}
The $t_{2g}$ orbital Hamiltonian in a ferromagnetic Mott insulator at zero temperature 
was examined in Ref.~\onlinecite{ko}. 
Through the analytical calculations,  
the two kinds of long-range orbital ordered states~\cite{ko2} were proposed. 
It was shown that these orders set in at only at zero temperature due to strong quantum fluctuation. 
However, the puzzling spin-orbital properties in YTiO$_3$ still remains an open issue. 
Even in a pure theoretical view point, finite-temperature ($T$) orbital states with large quantum fluctuation are unclear. 

In this paper, we study the $t_{2g}$ orbital system with ferromagnetic polarization. 
Our purpose is to examine finite-temperature orbital state in the idealized $t_{2g}$ orbital model by using an unbiased method. 
We employ the quantum Monte-Carlo (QMC) simulation with the loop algorithm. 
Down to at least around $T \sim 0.3J$ with the coupling constant $J$, indications of the conventional staggered-type order, the off-diagonal orbital order and the orbital dimerized state are not conspicuous. 
Instead, the orbital gap-like behaviors are found in the uniform orbital susceptibility below around 0.8$J$. 

We start from the Hamiltonian for the $(t_{2g})^1$ system 
with ferromagnetic polarization in an ideal Perovskite lattice. 
The generalized Hubbard Hamiltonian in a simple-cubic lattice 
with the triply-degenerate $t_{2g}$ orbitals is given as 
\begin{eqnarray}
{\cal H}_{0}&=&
\sum_{\langle ij \rangle \xi \xi' s} 
\left ( t_{ij}^{\xi \xi'} d_{i \xi s}^\dagger d_{j \xi' s}^{} +{\rm H.c.} \right )  
+U \sum_{i \ \xi} n_{i \xi \uparrow} n_{i \xi \downarrow} 
\nonumber \\
&+&\frac{1}{2} U'\sum_{i \ \xi \ne \xi'} n_{i \xi} n_{i \xi'}
+\frac{1}{2}K \sum_{i \ \xi \ne  \xi' s s'} 
d_{i \xi s}^\dagger d_{i \xi' s'}^\dagger d_{i \xi s'}^{} d_{i \xi' s}^{}
\nonumber \\
&+&I \sum_{i \ \xi \ne \xi'} d_{i \xi \uparrow}^\dagger d_{i \xi \downarrow}^\dagger 
d_{i \xi' \downarrow}^{} d_{i \xi' \uparrow}^{}. 
\label{eq:hubbard}
\end{eqnarray} 
We define the electron annihilation operator $d_{i \xi \sigma}$ 
with orbital $\xi=(\alpha, \beta, \gamma)$, 
spin $s=(\uparrow, \downarrow)$ at site $i$, and  
the electron transfer integral $t_{ij}^{\xi \xi'}$. 
A pair of the nearest neighboring (NN) sites is denoted by $\langle ij \rangle$. 
The intra-orbital Coulomb interaction $U$, the inter-orbital one $U'$, 
the Hund coupling $K$, and the pair-hopping $I$ are introduced. 
In an atomic limit, there is the relations $U=U'+2I$ and $K=I$. 
We also introduce a number operator, $n_{i \xi}=\sum_{s} n_{i \xi s}=\sum_s d^\dagger_{i \xi s} d_{i \xi s}$,  
which has a conservation $\sum_\xi n_{i \xi}=1$.   
When the electron transfer-integrals through the O $2p_{\pi}$ orbitals are only taken into account,  
the transfer integral is written as a simple form 
$t_{ij}^{\xi \xi'}=t \delta_{\xi \xi'} ( \delta_{a_l \xi}+\delta_{b_l \xi} ) $.  
A subscript $l(=x, y, z)$ indicates a direction connecting $i$ and $j$. 
We define two ``active" orbitals, $a_l$ and $b_l$, with a finite transfer-integral, 
and an ``inactive" one, $c_l$, with no transfer-integral.  
That is, 
$(a_x, b_x, c_x)=(\beta, \gamma, \alpha)$, 
$(a_y, b_y, c_y)=(\gamma, \alpha, \beta)$, and 
$(a_z, b_z, c_z)=(\alpha, \beta, \gamma)$. 
It is convenient to introduce the pseudo-spin (PS) operator for the active orbitals 
${\bf T}^l_i=\frac{1}{2}
\sum_{ s \ \xi \xi'=(a_l b_l) } d_{i \xi s }^\dagger {\bf \sigma}_{\xi \xi'} d_{i \xi' s}^{} $
with the Pauli matrices ${\bf \sigma}$.   
The effective Hamiltonian for the $(t_{2g})^1$ system which 
has strong on-site Coulomb interactions 
is obtained by the perturbational processes.~\cite{kugel82,ishihara02,ko}
The virtual intermediate states are classified by the irreducible representations for 
the $(t_{2g})^2$ states, i.e. ${}^1A_1$, ${}^1T_2$, ${}^1E$ and ${}^3T_1$. 
The lowest energy state is ${}^3 T_1$ with the energy of $U'-K$. 
We assume spins are saturated in ferromagnetic phase 
where ${}^3T_1$ is only the relevant intermediate state. 
The Hamiltonian studied in the present paper is 
\begin{eqnarray}
{\cal H}&=&-2J \sum_{\langle ij \rangle} \biggl [
n_{i a_l}n_{j b_l}+n_{i b_l}n_{j a_l}
\nonumber \\
&-&\left ( 
 d_{i a_l}^\dagger d_{i b_l}^{} d_{j b_l}^\dagger d_{j a_l}^{} 
+d_{i b_l}^\dagger d_{i a_l}^{} d_{j a_l}^\dagger d_{j b_l}^{}
 \right ) \nonumber \\
&+& \frac{1}{2} \left \{ n_{i c_l} \left ( n_{j a_l}+n_{j b_l} \right )
+\left (n_{i a_l}+n_{i b_l} \right ) n_{j c_l} \right \}
\biggr ] ,
\label{eq:hamiltonian0}
\end{eqnarray}
where the exchange constant is $J=t^2/(U'-K)$. 
This Hamiltonian is rewritten by using the PS operator as 
${\cal H}=4J \sum_{\langle ij \rangle} 
\{ {\bf T}^l_i \cdot {\bf T}^l_j  +  ( n_{i c_l} n_{j c_l}-1  )/4 \} $, 
and is the same Hamiltonian studied in Refs.~\onlinecite{ishihara02} and \onlinecite{ko}. 
In Ref.~\onlinecite{ko}, the authors proposed the two orbital orders 
where the wave-functions are the linear combinations of the $(d_\alpha, d_{\beta}, d_{\gamma})$ 
orbitals with equal weight,~\cite{ko2} which are called the off-diagonal type orbital orders in this paper. 

Before showing the numerical results, we remark some rigorous results for the orbital order. 
There is a novel symmetry in this Hamiltonian: 
in each plane perpendicular to a direction $l$, 
an electron number with the inactive orbital $c_l$, $\sum_{i}^\prime n_{i c_l}$, is conserved.~\cite{ishihara02,harris03,ko} 
A symbol $\sum^\prime_i$ implies a sum of sites in a plane being perpendicular to $l$. 
The Hamiltonian is invariant under the two-dimensional $U(1)$ gauge transformation denoted by 
$U(\theta)=\exp ( -i \theta \sum_i^\prime n_{i c_l} )$. 
By following the generalized Elitzur's theorem,~\cite{nussinov05,elitzur75} 
the physical quantities which are not invariant under this transformation 
has a vanishing mean value. 
This is proven from a combination of a theorem about a gauge transformation in a reduced dimension 
and the Mermin-Wagner theorem.~\cite{nussinov05,mermin66} 
By using the theorem, we prove that the long-range order for the following  
orbital operator does not appear at finite temperature, 
\begin{equation}
O_{\bf p}^l=
\sum_i e^{i{\bf p} \cdot {\bf r}_i } 
\left (
C_a d_{i c_l}^\dagger  d_{i b_l}^{} 
+C_b d_{i a_l}^\dagger  d_{i c_l}^{} +{\rm H.c.} 
\right ) , 
\label{eq:op}
\end{equation}
for $l=(x, y, z)$
with numerical constants $C_a$ and $C_b$.  
That is to say, $\langle O_{\bf p}^l \rangle=0$ at finite temperature. 

To analyze the Hamiltonian in finite temperature by using an unbiased method, we employ the QMC method with the loop algorithm.~\cite{evertz03,kawashima04} 
There is no negative-sign problem.   
Simulations are performed on $L^3$ cubic lattices with the periodic-boundary condition. 
The Suzuki-Trotter decomposition with the Trotter number $n$ is adopted. 
To check efficiency of the present QMC simulation, 
we calculate the auto-correlation time~\cite{kawashima94} 
for the parameter $L^{-3} \sum_i T_{iz}^x $. 
The auto-correlation time grows up monotonically with decreasing temperature, 
and reaches around 750MC steps at $T/J=0.3$. 
Thus, 
we adopt, in the following simulation, 2$\times 10^3$MC steps for thermalization, and 10$^4$MC steps for measurements. We estimate averages and errors for the physical quantities in 10 independent simulations. 
System size is taken to be $L=6 - 12$ in a cubic lattice. 
Numerical data obtained in the Trotter numbers $n$=10, 20 and 30 are extrapolated. 

\begin{figure}
\includegraphics[width=\columnwidth]{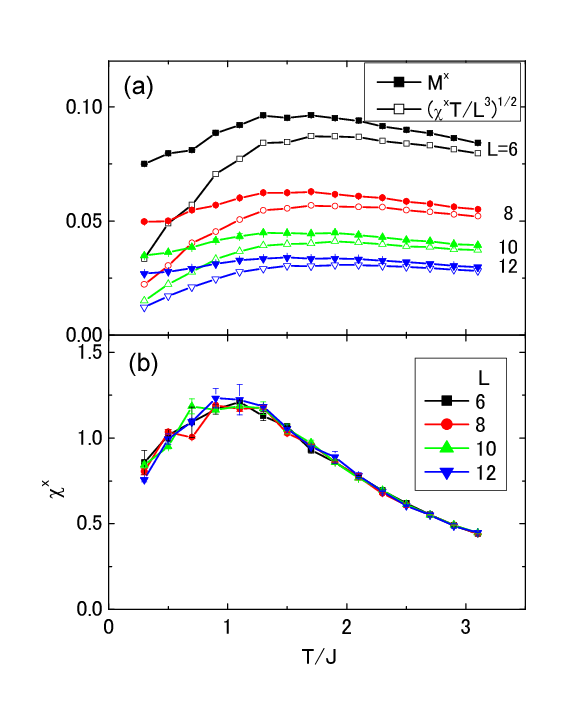}
\caption{  
(a) Staggered correlation function $M^x$ for the orbital PS operator $T_{i z}^x$. 
(b) Staggered orbital susceptibility $\chi^x$. 
For comparison, $\sqrt{\chi^x T/L^3}$ is also plotted in (a).}
\label{fig:mxchix}
\end{figure}

First, we show results for the staggered-type orbital order. 
We introduce the staggered correlation function along direction $l$,  
\begin{equation}
M^{l 2}  =\frac{4}{6n(L^3)^2} \left \langle \sum_\tau
\Bigl \{   \sum_i (-1)^i  T_{i z }^l(\tau)     \Bigr \}^2  
\right \rangle , 
\label{eq:ml}
\end{equation}
and the staggered orbital susceptibility, 
\begin{equation}
\chi^l=\frac{4}{NT}
\biggl \langle 
\Bigl \{ \sum_\tau \sum_{i} (-1)^i T_{i z}^{l } (\tau)  \Bigr \}^2 \biggr \rangle ,
\label{eq:chi}
\end{equation}
where $N=6nL^3$, $\tau$ is an imaginary time, $T^l_{i z}(\tau)$ is the $z$ component of the PS operator at $\tau$, 
and $\langle \cdots \rangle $ implies the QMC average. 
In the classical limit, $M^l$ should be identical to $\sqrt{\chi^l T /L^{3}}$.  
Numerical results of $M^x$ are presented in Fig.~\ref{fig:mxchix}(a). 
We have checked that three $M^l$ for $l=(x, y, z)$ coincide with each other within numerical errors. 
In $L=6$ case, $M^x$ shows a broad peak around $T/J=1.5$. 
This peak is, however, smeared out with increasing $L$. 
All obtained values are less than 10$\%$ of the maximum value, i.e. one. 
For comparison, $ \sqrt{\chi^x T /L^{3}}$ is also plotted in the same figure. 
In high temperatures, $M^x$ and $\sqrt{\chi^x T/L^{3}}$  merge together for all $L$ as expected.  
A deviation between the two is remarkable below $T/J \sim 1$ where 
the quantum effect starts to be effective. 
In both the two quantities, there are no divergent behaviors at $T=0$.  
As shown in Fig.~\ref{fig:mxchix}(b), numerical data of $\chi^x$ for all system size $L$ 
are scaled by a single curve in a whole temperature range. 
This implies that the correlation function 
$\sum_{i j \tau \tau'} (-1)^{i+j}  \langle   T_{iz}^l(\tau) T_{j z}^{l } (\tau')   \rangle $ in Eq.~(\ref{eq:chi}) 
is of the order of $L^3$. 
That is to say, this correlation is rather short ranged down to at least 0.3$J$. 
The obtained $\chi^x$ remains to be finite down to the lowest temperature. 

\begin{figure}[t]
\includegraphics[width=\columnwidth]{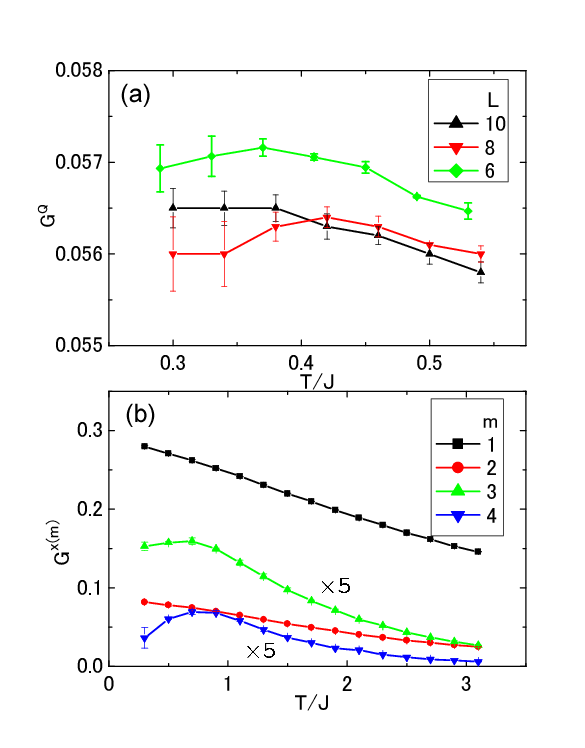}
\caption{  
(a) Correlation function $G^Q$ for the off-diagonal orbital operator $Q_i$. 
(b) Correlation function $G^{l(m)}$ for the orbital PS operator $T_{iz}^x$. 
System size in (b) is taken to be $L=10$. 
Results in $G^{x(3)}$ and $G^{x(4)}$ are multiplied by five.  
}
\label{fig:qq}
\end{figure}
This decrease in $M^l$ and $\chi^l$ below $T/J \sim 1$   
may be explained by a scenario that 
the orbital order of the off-diagonal operator proposed in Ref.~\onlinecite{ko} is developed in low temperatures; 
in both the two wave-functions claimed in Ref.~\onlinecite{ko},  
the correlation function $\langle T_{i z}^l T_{jz}^l \rangle$ for every 
NN pairs of sites $i$ and $j$ vanishes. 
To examine this possibility, 
we calculate the physical quantities for the off-diagonal orbital operator. 
In general, thermal average of the two-point function for an off-diagonal operator may be calculated by utilizing the improved estimator in the loop algorithm.~\cite{evertz03,kawashima04} 
In the present QMC simulation, however, 
this method is not realistic. 
This is because 
a value of the improved estimator depends on a shape of the loop, 
since the orbital interaction explicitly depends on the bond direction.  
Instead, the NN correlation functions for the off-diagonal operators 
are able to be calculated, when this is included in the Hamiltonian explicitly. 
We calculate the following correlation function for the off-diagonal operator 
$Q_i \equiv \sum_{l=(x,y,z)} T_{i x}^l$ defined by 
\begin{equation}
G^{Q}=\frac{1}{zN} \biggl \langle \sum_{\tau} \sum_{\langle ij \rangle} Q_i(\tau) Q_j(\tau) \biggr \rangle , 
\end{equation}
where $z=6$. 
For the orbital ordered states proposed in Ref.~\onlinecite{ko}, 
$G^Q=0.22$ and 0.16 for the type-I and type-II orders.  
Numerical results of $G^Q$ presented in Fig.~\ref{fig:qq}(a)  
show weak size dependence and do not show remarkable increasing with decreasing temperature. 
The obtained values are about 35$\%$ of the values expected from the ideal off-diagonal orbital orders.  

Further information for the off-diagonal orbital order is obtained by 
the staggered correlation as a function of distance: 
\begin{equation}
G^{l (m)}=\frac{4}{N} 
\bigg \langle 
\sum_\tau \sum_{\langle i j \rangle_l}' (-1)^{i+j} T^l_{i z}(\tau) T^l_{j z}(\tau) 
\bigg \rangle , 
\end{equation}
where $\sum_{\langle ij \rangle_l}^\prime$ represents a sum for the $m$-th NN sites 
$i$ and $j$ along direction $l$. 
We calculate $G^{l (m)}$'s for $l=x$ up to $m=4$ [see Fig.~\ref{fig:qq}(b)]. 
With decreasing temperature, 
$G^{x (1)}$ and $G^{x(2)}$ monotonically increase even below $T/J \sim 1$ where 
$M^l$ and $\chi^l$ in Fig.~\ref{fig:mxchix} start to decrease. 
On the other hand, reductions are seen in $G^{x (3)}$ and 
$G^{x (4)}$ around $T/J=0.8$. 
That is, the decreases in $M^l$ and $\chi^l$ 
are attributed to those in the long-range correlations of $T_{iz}^l$, 
and short-range correlations still remain to grow up. 
The obtained result also suggests that the decreases in $M^l$ and $\chi^l$ 
are not attributed to development of the off-diagonal orbital correlation; 
when the correlation for the off-diagonal operator $Q_i$ grows up at low temperature, 
reduction in $G^{l (m)}$ should be remarkable in short range. 
That is, $G^{l(1)}$ and $G^{l(2)}$, instead of $G^{l(3)}$ and $G^{l(4)}$, are expected to be reduced. 
This expectation is in contrast to the numerical results in Fig.~\ref{fig:qq}(b). 

\begin{figure}[t]
\includegraphics[width=\columnwidth]{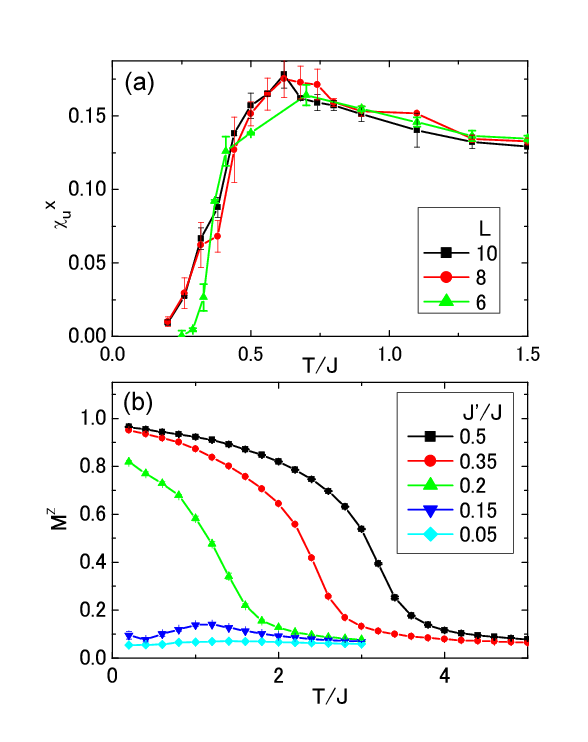}
\caption{(a) Uniform susceptibility for the orbital PS operator $T_{iz}^x$.
(b) Correlation function $M^z$ calculated in the Hamiltonian ${\cal H}+{\cal H}'$. 
System size and the Trotter number in (b) are chosen to be $L=8$ and $n=10$, respectively. 
}
\label{fig:uniform}
\end{figure}
Development of the short-range orbital correlation shown above suggests a kind of the valence-bond state 
in low-dimensional quantum magnets.~\cite{troyer96} 
However, we are able to exclude a simple orbital dimer state. 
We calculate the dimer order parameter defined as  
\begin{equation}
D^{l2}=
\frac{16}{6n(L^3)^2}
\biggl \langle \sum_\tau 
\biggl \{  \sum_{i} T_{i z}^l (\tau) T_{i+{\bf e}_l z }^l (\tau) \biggr \}^2   
\biggr \rangle , 
\end{equation}
with a unit vector ${\bf e}_l$ along direction $l$. 
Results of $D^l$ show monotonic decreases with decreasing temperature and increasing the system size (not presented in figure). 

The following results indicate that the low-temperature orbital state shows 
a gap-like feature.  
The calculated uniform orbital susceptibility defined by 
\begin{equation}
\chi^l_u=\frac{4}{NT}
\biggl \langle 
\Bigl \{ \sum_\tau \sum_i T_{i z}^{l } (\tau)  \Bigr \}^2 \biggr \rangle , 
\label{eq:chiu}
\end{equation}
[see in Fig.~\ref{fig:uniform}(a)] starts to decrease around $T/J=0.7$ 
and tends to vanish at low temperatures. 
Size dependence is seen below $T/J \sim 0.3$, but a global gap-like feature in $\chi_u^l$ is not sensitive to the system size. 
The temperature, where $\chi_u^l$ starts to decrease, i.e. $T \sim 0.7J$, almost coincides to the temperature where the staggered susceptibility $\chi^x$ and 
the short-range correlation function $G^{x(3)}$ and $G^{x(4)}$ have broad peaks, as shown in Figs.~\ref{fig:mxchix}(b) and \ref{fig:qq}(b). 
This gap-like feature is also confirmed by the following calculation. 
We add the Ising-like interaction term into the Hamiltonian Eq.~(\ref{eq:hamiltonian0}):  
\begin{eqnarray}
{\cal H}'=2J' \sum_{\langle ij \rangle} \left ( 
4T_{iz}^z  T_{jz}^z -n_{i \alpha} n_{j \alpha}-n_{i \beta} n_{j \beta}  
\right ) , 
\label{eq:h2}
\end{eqnarray}
where $J'$ is a positive coupling constant. 
This term promotes the staggered-type orbital order characterized by $M^z$. 
Results are shown in Fig.~\ref{fig:uniform}(b) for several values of $J'/J$.  
A threshold value for $J'/J$ seems to exist; 
an increase of $M^z$ in low temperatures is not seen for $J'/J=0.05$ and 0.15. 
This is consistent with finite values of $\chi^x$ at low temperatures shown in Fig.~\ref{fig:mxchix}(b). 
These results indicate a possibility that a quantum gapped state is broken by the Ising-type interaction 
as well known in the quantum critical issue.  

In summary, we present a numerical study of the finite-temperature $t_{2g}$ orbital state in a ferromagnetic Mott insulator. 
Remarkable developments are not seen in the staggered orbital correlation and the NN correlation for the off-diagonal orbital operator at least down to $T/J \sim 0.3$. 
Short-range correlation remains to grow up in low temperatures, but a possibility of the dimerized state is excluded. 
It is found that the uniform orbital susceptibility shows a gap-like feature below $T/J \sim 0.7$. 
This result is supported from a calculation in the model where the Ising-like interaction is added. 
One possible scenario for the low temperature orbital state is a short-range singlet state where singlet pairs are randomly distributed in a crystal, and/or rearrangement of pairs occurs dynamically.  
A singlet wave function between sites $i$ and $j$ are  
$(|d_{i a_l }  d_{j b_l} \rangle-|d_{i b_l} d_{j a_l} \rangle )/\sqrt{2}$ where concerning orbitals depend on the bond direction $l$, i.e. a directional singlet pair. 
To clarify a low temperature orbital state in more detail, 
QMC simulations in large cluster sizes with efficient algorithms are 
necessary. A variational-type approach based on the above directional singlet assumption may be also helpful.

The authors would like to thank K.~Harada, 
M. Matsumoto, H. Matsueda and Z.~Nussinov for their valuable discussions. 
The authors also thank T.~Watanabe and J.~Nasu for their critical reading of the manuscript.  
This work was supported by JSPS KAKENHI (16104005) and 
TOKUTEI (18044001, 19052001, 19014003) from MEXT, 
NAREGI, and CREST.

$^{\ast}$ Present address: The Institute for Solid State Physics, University of Tokyo, Kashiwa, Chiba 277-8581, Japan.


\begin{references}
\bibitem{ko2}
The two off-diagonal type orbital orders proposed in Ref.~\onlinecite{ko}  
consist of four orbital sublattices in a simple cubic lattice.  
The wave functions are given by 
$
| \psi_i^{\rm I}  \rangle= 
( C_i^\alpha  | d_{yz}  \rangle + C_i^\beta  | d_{zx}  \rangle 
+ C_i^\gamma  | d_{xy}  \rangle )/\sqrt{3} , 
$
for the type I, and 
$
| \psi_i^{\rm II}  \rangle= 
( C_i^\alpha e^{i 2\pi/3} | d_{yz} \rangle + C_i^\beta e^{-i 2\pi/3}
| d_{zx}  \rangle + C_i^\gamma  | d_{xy}  \rangle    )/\sqrt{3} , 
$
for the type II, in the classical limit. 
Subscript $i(=1-4)$ denotes the four sublattices, and 
coefficients are 
$(C_i^\alpha, C_i^\beta, C_i^\gamma)=(1, 1,1 )$ for $i=1$, $(-1, -1, 1)$ for $i=2$, 
$(-1, 1,-1)$ for $i=3$, and $(1, -1, -1)$ for $i=4$.
\end{references}
\end{document}